\documentstyle[hep93]{article}
\begin{document}
\title{Generalized Deformed Oscillators and Algebras}
\author{Dennis Bonatsos $^{*}$,
C. Daskaloyannis $^\dagger$,
P. Kolokotronis $^*$\\
   {\em $^*$ Institute of Nuclear Physics, NCSR ``Demokritos''}\\
   {\em GR-15310 Aghia Paraskevi, Attiki, Greece }\\
 {\em $^\dagger$ Department of Theoretical Physics, Aristotle University of
            Thessaloniki}\\
       {\em GR-54006 Thessaloniki, Greece} }

\abstract{
The generalized deformed oscillator schemes introduced as unified frameworks
of various deformed oscillators are proved to be equivalent, their unified
representation leading to a correspondence between the deformed oscillator and
the N=2 supersymmetric quantum mechanics (SUSY-QM) scheme. In addition, several
physical systems (two identical particles in two dimensions, iso\-tro\-pic
oscillator and Kepler system in a 2-dim curved space) and mathematical
structures (quadratic algebra QH(3), finite W algebra $\bar {\rm W}_0$) are
shown to posses the structure of a generalized deformed su(2) algebra, the
representation theory of which is known. Furthermore, the generalized deformed
parafermionic oscillator is identified with the algebra of several physical
systems (iso\-tro\-pic oscillator and Kepler system in 2-dim curved space,
Fokas--Lagerstrom, Smorodinsky--Winternitz and Holt potentials) and
mathematical constructions (generalized deformed su(2) algebra, finite W
algebras $\bar {\rm W}_0$ and W$_3^{(2)}$). The fact that the Holt potential
is characterized by the W$_3^{(2)}$ symmetry is obtained as a by-product.
}

\maketitle
\section{Introduction}

Quantum algebras (quantum groups) \cite{Dri798,Jim247},  are nonlinear
generalizations of the usual Lie algebras to which they reduce for
appropriate values of the deformation parameter(s). From the mathematical
point of view they are Hopf algebras \cite{Abe77}.
Their use in physics became popular with the introduction of the
$q$-deformed harmonic oscillator \cite{Bie873,Mac4581,SunFu} as a tool
for providing a boson realization of the quantum algebra su$_q$(2),
although similar mathematical structures had already been known
\cite{AC524,Kur111}. Initially used for solving the quantum Yang--Baxter
equation \cite{Jim10}, quantum algebras have subsequently found
applications in several branches of physics, as, for example, in
the description of spin chains, squeezed states, rotational and vibrational
nuclear and molecular spectra, and in conformal field theories.
By now several kinds of generalized deformed oscillators have been introduced
\cite{Das789,DY4157,Da2261}
and unification schemes for them have been suggested \cite{PLB307}.

Furthermore, generalized deformed su(2) algebras have been introduced
\cite{BDK871,Pan5065}
in a way that their representation theory remains as close
as possible to the usual su(2) one. It will be shown here that several
physical systems (two identical particles in two dimensions \cite{LM3649},
isotropic oscillator and Kepler system in a 2-dim curved space
\cite{Higgs309},
as well as the quadratic algebra QH(3) \cite{GLZ217} and the finite W
algebra $\bar {\rm W}_0$ \cite{Bow945}
can be accommodated within this scheme. The advantage this
unification offers is that the representation theory of the generalized
deformed su(2) algebras is known \cite{BDK871}.

In addition, generalized parafermionic oscillators have been introduced
\cite{Que245}, in analogy to generalized deformed oscillators.
It will be shown here that several physical systems (isotropic oscillator
and Kepler system in a 2-dim curved space \cite{Higgs309}, Fokas--Lagerstrom
potential \cite{FL325},
Smorodinsky--Winternitz potential \cite{WS444}, Holt potential
\cite{Holt1037}) and mathematical constructions (generalized deformed su(2)
algebra \cite{BDK871,Pan5065}, finite W algebras $\bar {\rm W}_0$
\cite{Bow945} and W$_3^{(2)}$ \cite{Tjin60,Tjin485,Tjin93,Tjin161}) can be
accommodated within this framework.
As a by-product the fact that the Holt potential is characterized by the
W$_3^{(2)}$ symmetry occurs \cite{BDK2197}.

In section 2 the various deformed oscillators will be put in a common
mathematical framework, while in section 3 various unification schemes will
be proved to be equivalent.
In section 4 the relation of the deformed oscillator to the N=2
supersymmetric quantum mechanics (SUSY-QM) scheme will be pointed out.
 In section 5 the  cases related  to the generalized
deformed su(2) algebra will be studied, while in section 6 the systems
 related to generalized deformed
parafermionic oscillators will be  considered.
Section 7 will contain
discussion of the present results and plans for further work.

\section{The generalized deformed oscillator}

By now
many kinds of deformed oscillators have been introduced in the literature.
All of them can be accommodated within the common mathematical framework
of the {\sl generalized deformed oscillator} \cite{Das789,DY4157,Da2261},
which is defined
as the algebra generated by the operators $\{1, a, a^\dagger, N\}$ and the
{\sl structure function} $\Phi(x)$, satisfying the relations
$$ [a, N]=a, \qquad [a^\dagger, N]=-a^\dagger, \eqno(2.1)$$
$$ a^\dagger a =\Phi(N) =[N], \qquad aa^\dagger = \Phi(N+1) =[N+1],
\eqno(2.2)$$
where $\Phi(x)$ is a positive analytic function with $\Phi(0)=0$ and $N$ is
the number operator.
{}From eq. (2.2) we conclude that
$$N=\Phi^{-1} (a^\dagger a),\eqno(2.3)$$
and that the following commutation and anticommutation relations are
obviously satisfied:
$$ [a, a^\dagger]=[N+1]-[N], \qquad \{a,a^\dagger\}=[N+1]+[N].\eqno(2.4)$$
The {\sl structure function} $\Phi(x)$ is characteristic to the deformation
scheme. In table 1 the structure functions corresponding to different
deformed oscillators are given. They will be further discussed at the end of
this section.

It  can be proved that
the generalized deformed algebras possess a Fock space of
 eigenvectors
$|0>$, $|1>$, $\ldots$, $|n>$, $\ldots$
of the number operator $N$
$$N|n>=n|n>,\quad <n|m>=\delta_{nm}, \eqno(2.5) $$
if the {\sl vacuum state} $|0>$ satisfies the following relation:
$$ a|0>=0. \eqno(2.6)$$
 These eigenvectors are generated by the formula:
 $$ \vert n >= {1 \over \sqrt{ [n]!}} {\left( a^\dagger \right)}^n \vert 0 >,
\eqno(2.7) $$
where
 $$[n]!=\prod_{k=1}^n [k]= \prod_{k=1}^n \Phi(k). \eqno(2.8) $$
The generators  $a^\dagger$ and $a$ are the creation and
annihilation operators of this deformed oscillator algebra:
$$a\vert n> = \sqrt{[n]} a\vert n-1>,\qquad
 a^\dagger \vert n> = \sqrt{[n+1]} a\vert n+1>. \eqno(2.9) $$

These eigenvectors are also eigenvectors of the energy operator
$$H={\hbar \omega \over 2} (aa^\dagger +a^\dagger a), \eqno(2.10)$$
corresponding to the eigenvalues
$$ E(n)= {\hbar\omega \over 2} (\Phi(n)+\Phi(n+1)) =
{\hbar \omega \over 2} ([n]+[n+1]).\eqno(2.11)$$

For $$\Phi(n)=n  \eqno(2.12)$$
one obtains the results for the ordinary harmonic oscillator.
For $$\Phi(n)= {q^n -q^{-n}\over q-q^{-1}}=[n]_q  \eqno(2.13)$$
one has the results for the $q$-deformed harmonic oscillator
\cite{Bie873,Mac4581,SunFu},
while the choice
$$ \Phi(n) ={Q^n-1\over Q-1}= [n]_Q  \eqno(2.14)$$
leads to the results of the $Q$-deformed harmonic oscillator
\cite{AC524,Kur111}. Many more
cases are shown in table 1, on which the following comments apply:


\begin{table}[bt]
\begin{center}
\caption{ Structure functions of special deformation schemes}
\bigskip
\begin{tabular}{|c c p{2.0 in}|}
\hline
\ & $\Phi(x)$ & Reference \\
\hline\hline
\romannumeral 1 & $x$ &
harmonic  oscillator, bosonic algebra\\[0.05in]
\romannumeral 2&  ${ {q^x- q^{-x} }  \over {q- q^{-1} } } $ &
$q$-deformed harmonic oscillator \cite{Bie873,Mac4581,SunFu}
\\[0.05in]
\romannumeral 3& ${ {q^x- 1 } \over {q- 1 } } $ & Arik--Coon,
Kuryshkin, or $Q$-deformed oscillator \cite{AC524,Kur111} \\[0.05in]
\romannumeral 4&  ${ {q^x- p^{-x} }  \over {q- p^{-1} } } $ &
2-parameter deformed oscillator \cite{BJM820,JBM775,CJ711}
\\[0.05in]
\romannumeral 5& $x(p+1-x)$ & parafermionic oscillator
 \cite{OK82} \\[0.05in]
\romannumeral 6& $ { \sinh (\tau x) \sinh (\tau (p+1-x) )}\over
{ \sinh^2(\tau) } $ & $q$-deformed parafermionic oscillator
\cite{FV1019,OKK591} \\[0.05in]
\romannumeral 7& $x\cos^2(\pi x/2) + (x+p-1)\sin^2(\pi x /2)  $&
parabosonic oscillator \cite{OK82} \\[0.05in]
\romannumeral 8&
$\begin{array}{c}
\frac{\sinh(\tau x)}{\sinh(\tau)}
\frac{\cosh(\tau (x+2N_0-1))}{\cosh(\tau)} \cos^2 (\pi x/2) +\\
+ \frac{\sinh(\tau (x+2N_0-1))}{\sinh(\tau)}
\frac{\cosh(\tau x)}{\cosh(\tau)} \sin^2 (\pi x/2)
\end{array}$
 & $q$-deformed parabosonic oscillator
 \cite{FV1019,OKK591} \\[0.05in]
\romannumeral 9 &
$\sin^2 {\pi x/2}$ & fermionic algebra \cite{JBSZ123} \\[0.05in]
\romannumeral 10 & $ q^{x-1} \sin^2 {\pi x/2}$ &
$q$-deformed fermionic algebra \cite{Hay129,CK72,FSS179,Gang819,SCi219,PV613}
 \\[0.05in]
\romannumeral 11&
$\frac{1-(-q)^x}{1+q}$ & generalized $q$-deformed fermionic algebra
 \cite{VPJ335} \\[0.05in]
\romannumeral 12& $x^n$ & \cite{Das789} \\[0.05in]
\romannumeral 13& ${ {sn(\tau x)} \over {sn(\tau )} }$ & \cite{Das789}
 \\[0.05in]
\hline
\end{tabular}
\end{center}
\end{table}

i) Two-parameter deformed oscillators have been introduced
\cite{BJM820,JBM775,CJ711}, in analogy to the one-parameter deformed
oscillators.

ii) Parafermionic oscillators \cite{OK82} of order $p$ represent particles
of which the maximum number which can occupy the same state is $p$.
Parabosonic oscillators \cite{OK82} can also be introduced.

iii) $q$-deformed versions of the parafermionic and parabosonic oscillators
have also been introduced \cite{FV1019,OKK591}.

iv) $q$-deformed versions of the fermionic algebra \cite{JBSZ123} have also
been introduced \cite{Hay129,CK72,FSS179,Gang819,SCi219,PV613}, as well as
$q$-deformed versions of generalized $q$-deformed fermionic algebras
\cite{VPJ335}. It has been proved, however, that $q$-deformed fermions are
fully equivalent to the ordinary fermions \cite{BD1589,JX891,Kong92}.

\section{Equivalence of unification schemes of deformed oscillators}

Using the definition (eqs (2.1), (2.2)) of
the deformed algebra we can find the relation between the present
formalism and the various deformation frameworks introduced recently.
For each deformation framework the structure function
$\Phi(x)$ can be determined. Thus each deformation scheme is
equivalent to the unified treatment expressed by eqs (2.1) and (2.2).

\subsection{The Beckers--Debergh unification scheme}

 Beckers and Debergh proposed recently \cite{BD1277} a unification scheme
based on the following set of relations:
$$ a a^\dagger + g(q) a^\dagger a = |[N+1]|+g(q)|[N]|, \eqno(3.1)$$
$$ c^\dagger|n>  = \sqrt{|[n+1]|} |n+1>, \eqno(3.2)$$
$$ c|n>  = \sqrt{|[n]|}|n-1>, \eqno(3.3)$$
where $g(q)$ is an ordinary function of the parameter $q$ and
the bracket $[n]$ is a function of $n$. Obviously the
correspondence between the Beckers-Debergh formalism and the
algebra defined by eqs (2.1), (2.2) is
given by defining the structure function $\Phi(x)$ as follows:
$$ \Phi(x)=|[x]| . \eqno(3.4)$$
In \cite{BD1277} eq. (2.2) is  not used explicitly, eq. (3.1) being
used instead. If one considers eq. (2.2) as the fundamental one,
then eqs (3.1)-(3.3) are satisfied, simplifying
considerably the demonstrations of the theorems.

\subsection{ The Odaka--Kishi--Kamefuchi unification scheme}

 Odaka--Kishi--Kamefuchi \cite{OKK591} proposed an algebra generated by
the following relations:
$$ \big[ a, {\cal N}\big] =a,\quad \big[ a^\dagger, {\cal N} \big]=
-a^\dagger,
\eqno(3.5)$$
$$ {\big[ a^\dagger, a \big]}_\alpha = a^\dagger a + \alpha a a^\dagger =
G( {\cal N}),\eqno(3.6)$$
where the spectrum of the operator $\cal N $ is given by:
$$ {\cal N} |n> = \left( n + N_0 \right) |n>.\eqno(3.7)$$
After a little algebra Odaka--Kishi--Kamefuchi \cite{OKK591} define
the numbers $I_n$:
$$ I_n = \left( <n|a|n+1> \right)^2 =
\sum\limits_{m=0}^n (-1)^m \alpha ^{-(m+1)} G_{(n-m)}, \eqno(3.8)$$
where $G_n= <n|G({\cal N})|n>$.
Without difficulty we can find the correspondence between the
general definition (eqs (2.1), (2.2)) and the formalism of
the Odaka--Kishi--Kamefuchi (eqs (3.5)-(3.8)).
The number operator $N$ in eq. (2.2) is related
to the operator $\cal N$ by the relation:
$$N={\cal N} - N_0 = {\cal N}- <0|{\cal N}|0>, \eqno(3.9)$$
while the structure function $\Phi(x)$ is the solution of the
equation:
$$ \Phi (x)+\alpha \Phi (x+1) = G(x+N_0). \eqno(3.10)$$
The stucture function $\Phi(x)$ is also related to the numbers
$I_n$ in eq. (3.8) by:
$$ \Phi(n+1)= I_{n}. \eqno(3.11)$$

In many concrete cases eq. (3.11) can be used
directly for finding the correspondence between the
various formalisms. An example is the case of the
parabosonic oscillator \cite{OK82}. In eq. (6) of \cite{OKK591}
 the numbers $I_n$ are determined by:
$$ I_n= \cases {n+2 N_0 & for $n$=even \cr
                n+1     & for $n$=odd \cr} \eqno(3.12)$$
{}From eq. (3.11)  we find that
$$ \Phi(x)=x\cos^2(\pi x/2) +(x+2N_0-1) \sin^2 (\pi x/2), \eqno(3.13)$$
while in the case of the $q$-deformed version of the parabosons
(eq. (10) of \cite{OKK591}) the structure function is given by
$$ \Phi(x)
 = {\sinh(\tau x) \over \sinh(\tau)}
{\cosh(\tau (x+2N_0-1)) \over \cosh(\tau)} \cos^2 (\pi x/2)
 + {\sinh(\tau (x+2N_0-1)) \over \sinh(\tau)}
{\cosh(\tau x) \over \cosh(\tau)} \sin^2 (\pi x/2)  .\eqno(3.14)$$

\subsection{ The generalized deformed oscillator}

The method of the generalized deformed oscillator
 proposed by Daskaloyannis \cite{Das789,DY4157,Da2261}  is essentially
the same with the algebra generated by eqs (2.1) and (2.2). In \cite{Das789}
the initial assumption was that there is a function $g(x)$ such
that:
$$ a a^\dagger = g(a^\dagger a). \eqno(3.15)$$
This assumption is equivalent to the statement given
by eq. (2.2), the structure function $\Phi(x)$
being given by the solution of the functional equation:
$$ \Phi \left( 1 + \Phi^{-1}(x) \right) =g(x) .\eqno(3.16)$$

\subsection{ The bosonization scheme}

This method was initially proposed by Jannussis {\it et al.}
\cite{JBSZ123}.
 The algebra is defined by:
$$ a=F(N+1)b,\quad a^\dagger= b^\dagger F(N+1), \eqno(3.17)$$
where $\big\{ 1, b,b^\dagger,N \big\}$ is the usual oscillator algebra
$$ b^\dagger b=N, \quad b b^\dagger =N+1, \quad \big[ b,N \big]=b .
\eqno(3.18)$$
This theory is formally equivalent to the one proposed in this paper
by defining the structure function to be:
$$ \Phi(x)=xF(x). \eqno(3.19) $$

\subsection{ The generalized Q-deformed oscillator}

This algebra was
proposed by Brodimas {\it et al.} \cite{BJSZP177}. It is based on the
relation:
$$ aa^\dagger - Q a^\dagger a = f(N) . \eqno(3.20)$$
The corresponding structure function is:
$$ \Phi (x+1) - Q \Phi(x) = f(x).\eqno(3.21)$$

\section { Relation between  the generalized deformed oscillator and the
SUSY-QM}

The generation of the unified deformed oscillator schemes from
eq. (2.2)  shows the relation of this kind of
theories with the $N=2$ supersymmetric quantum mechanics (SUSY-QM)
\cite{LRB1383,Suk2917,Suk2937}.

Using the vocabulary of the SUSY-QM, we can
identify the operators $a^\dagger a$ and $aa^\dagger$ as being the partner
hamiltonians $H_+$ and $H_-$ of
the SUSY-QM.
 These partner hamiltonians
have common eigenstates $|n>$ corresponding to the
eigenvalues $\Phi(n)$ and $\Phi(n+1)$,
$$ H_+ |n>=E^+_n|n>, \qquad  E^+_n=\Phi(n) , \eqno(4.1)$$
$$ H_- |n>=E^-_n|n>, \qquad  E^-_n=\Phi(n+1). \eqno(4.2)$$
It is clear that $E^-_n=E^+_{n+1}$, while $\Phi (0) =0$
is an eigenvalue only of the Hamiltonian $H_+$ and not of
$H_-$. Thus for $H_+$ the state $|0>$ represents the ground state
with eigenvalue $E^+_0=0$, while no eigenstate with
zero eigenvalue exists for $H_-$ \cite{Suk2917,Suk2937}.  The ground state
eigenvalue of $H_-$ is non-zero.

The operators $a^\dagger$ and $a$ correspond to  the raising and lowering
operators of the SUSY-QM.
 The supersymmetric charges can be defined as:
$$ Q=\left(\matrix{ 0 & 0\cr a & 0 \cr}\right), \qquad
Q^+=\left( \matrix{ 0 & a^\dagger \cr 0 & 0 \cr}\right).\eqno(4.3)$$
The SUSY hamiltonian is defined by:
$$ H_S=\big\{ Q , Q^+ \big\} =
\left(\matrix{ H_+ & 0 \cr 0 & H_- \cr}
 \right) .\eqno(4.4)$$
It follows then that
$$ \big[Q,H_S \big]=\big[Q^+,H_S \big]=0, \qquad
Q^2= \left(Q^+\right)^2 =0 .\eqno(4.5)$$
Eqs (4.3)-(4.5) suggest that the unified schemes for deformed oscillators
are essentially similar to the SUSY-QM.

Starting from this point of view one can generate a
SUSY-QM using, for example, the $q$-deformed oscillator
of Biedenharn \cite{Bie873} and Macfarlane \cite{Mac4581}.
The present construction is an example of usual SUSY-QM,
different from the $q$-deformed N=2 SUSY-QM recently constructed
by Spiridonov \cite{Spi1241}.

 We have so far proved that the various unified deformation
schemes existing in the literature
 are equivalent
methods of creating a deformed oscillator. Their common root is
based on the fact that all these theories can be deducted from an operator
algebra $\big\{1,a,a^\dagger,N\big\}$ with the properties:
$$ \Big[ a , N \Big] = a, \qquad
 \left[ a^\dagger , N \right] = -a^\dagger , \eqno(4.6)$$
$$ a^\dagger a=\Phi(N)=[N], \qquad aa^\dagger=\Phi(N+1)=[N+1], \eqno(4.7)$$
where $\Phi(x)$ is a positive analytic {\it structure function} with
$\Phi(0)=0$. The above formulation shows the relation between  the
deformed oscillator formalism and the usual N=2 SUSY-QM. To every  deformed
oscillator corresponds an N=2  SUSY example.

An interesting problem is to examine
 which kind of multidimensional deformed
oscillator algebra could correspond to N$>$2 SUSY-QM. Another
interesting question is to study  if the correspondence

\centerline{deformed oscillator $\Longrightarrow$ N=2  SUSY }

\noindent
is invertible, i.e. if every N=2 SUSY algebra corresponds to a
deformed oscillator scheme.  It is also well known that one can find
SUSY algebras generated by superpotentials. The relation between
superpotentials and deformed oscillators should be examined.
{}From  the physical point of view, the implications of the present
findings on the recent effort of describing isospectral superdeformed
bands in neighbouring nuclei by SUSY-QM techniques \cite{ABCD2777} should
also be studied.

\section{ Generalized deformed su(2) algebras}

Generalized deformed su(2) algebras having representation theory similar
to that of the usual su(2) have been constructed in \cite{BDK871}. It has
been proved that it is possible to construct an algebra
$$ [J_0, J_{\pm}]=\pm J_{\pm}, \qquad [J_+,J_-]=\Phi(J_0(J_0+1))-
\Phi(J_0(J_0-1)),\eqno(5.1)$$
where $J_0$, $J_+$, $J_-$ are the generators of the algebra and
$\Phi(x)$ is any increasing entire function defined for $x\geq -1/4$.
Since this algebra is characterized by the function $\Phi$, we use for it
the symbol su$_{\Phi}$(2). The appropriate basis $|l,m>$ has the
properties
$$ J_0|l,m> = m |l,m>, \eqno(5.2)$$
$$ J_+|l,m> = \sqrt{\Phi(l(l+1))-\Phi(m(m+1))} |l,m+1>, \eqno(5.3)$$
$$ J_-|l,m> = \sqrt{\Phi(l(l+1))-\Phi(m(m-1))} |l, m-1>,\eqno(5.4) $$
where
$$ l=0, {1\over 2}, 1, {3\over 2}, 2, {5\over 2}, 3, \ldots,\eqno(5.5)$$
  and
$$ m= -l, -l+1, -l+2, \ldots, l-2, l-1, l.\eqno(5.6)$$
The Casimir operator is
$$ C= J_-J_+ +\Phi(J_0(J_0+1))=J_+J_-+\Phi(J_0(J_0-1)),
\eqno(5.7)$$
its eigenvalues indicated by
$$C |l,m> = \Phi(l(l+1)) |l,m>.\eqno(5.8)$$
The usual su(2) algebra is recovered for
$$ \Phi(x(x+1))= x(x+1),\eqno(5.9)$$
while the quantum algebra su$_q$(2)
$$ [J_0, J_{\pm}]=\pm J_{\pm}, \qquad [J_+, J_-]=[2 J_0]_q,\eqno(5.10) $$
occurs for
$$ \Phi(x(x+1))= [x]_q [x+1]_q,\eqno(5.11)$$
with $q$-numbers defined as
$$ [x]_q ={q^x-q^{-x}\over q-q^{-1}}.\eqno(5.12)$$

The su$_{\Phi}$(2) algebra occurs in several cases, in which the rhs of the
last equation in (5.1) is an odd function of $J_0$.

\subsection{ Two identical particles in two dimensions}

Let us consider the system of two identical particles in two
dimensions. For identical particles observables of the system have to be
invariant under exchange of particle indices.
A set of appropriate observables in this case is \cite{LM3649}
$$ u=(x_1)^2+(x_2)^2, \qquad v=(x_1)^2-(x_2)^2, \qquad w=2 x_1 x_2,
\eqno(5.13)$$
$$ U=(p_1)^2+(p_2)^2, \qquad V=(p_1)^2-(p_2)^2, \qquad W=2p_1 p_2,
\eqno(5.14)$$
$$ C_1= {1\over 4} (x_1 p_1+p_1 x_1), \qquad C_2={1\over 4} (x_2 p_2+
p_2 x_2), \qquad M= x_1 p_2+x_2 p_1,\eqno(5.15)$$
where the indices 1 and 2 indicate the two particles.
These observables are known to close an sp(4,R) algebra. A representation
of this algebra can be constructed \cite{LM3649,BRS425}
using one arbitrary constant
$\eta$ and three matrices $Q$, $R$, and $S$ satisfying the commutation
relations
$$ [S,Q]= -2iR, \qquad [S,R]=2iQ, \qquad [Q,R]= -8iS (\eta-2 S^2).
\eqno(5.16)$$
The explicit expressions of the generators of sp(4,R) in terms of $\eta$,
$S$, $Q$, $R$ are given in \cite{LM3649} and need not be repeated here.
Defining the operators
$$ X=Q-iR, \qquad Y=Q+iR, \qquad S_0= {S\over 2},\eqno(5.17)$$
one can see that the commutators of eq. (5.16) take the form
$$ [S_0,X]=X, \qquad [S_0,Y]=-Y, \qquad [X,Y]=32S_0 (\eta-8 (S_0)^2),
\eqno(5.18)$$
which is a deformed version of su(2).
It is clear that the algebra of eq. (5.18) is a special case of an
su$_{\Phi}$(2) algebra with structure function
$$ \Phi(J_0(J_0+1))= 16 \eta J_0(J_0+1) -64 (J_0(J_0+1))^2.\eqno(5.19)$$
The condition that $\Phi(x)$ has to be an increasing function of $x$
implies the restriction $x<\eta/8$.

\subsection{ Kepler problem in 2-dim curved space}

Studying the Kepler problem in a two-dimensional curved space with
constant curvature
$\lambda$ one finds the algebra \cite{Higgs309}
$$ [L, R_{\pm}]=\pm R_{\pm}, \qquad  [R_-, R_+]= F\left(L+{1\over 2}\right)-
F\left(L-{1\over 2}\right), \eqno(5.20)$$
where $$F(L)= \mu^2 + 2H L^2 -\lambda L^2\left(L^2-{1\over 4}\right).
\eqno(5.21)$$
It is then easy to see that
$$[R_+, R_-]= 2L\left(-2H+{\lambda\over 4}\right) + 4\lambda L^3,\eqno(5.22)$$
which corresponds to an su$_{\Phi}$(2) algebra with
$$ \Phi(J_0(J_0+1)) = \left(-2H+{\lambda\over 4}\right) J_0(J_0+1) +\lambda
(J_0(J_0+1))^2.\eqno(5.23)$$
For $\Phi(x)$ to be an increasing function, the condition
$$ \lambda x > H -{\lambda\over 8}\eqno(5.24)$$
has to be obeyed.

\subsection{ Isotropic oscillator in 2-dim curved space}

 In the case of the isotropic oscillator in a two-dimensional curved
space with constant curvature $\lambda$ one finds the algebra
\cite{Higgs309}
$$ [L,S_{\pm}] =\pm 2 S_{\pm}, \qquad [S_-, S_+]= G(L+1)-G(L-1),\eqno(5.25)$$
with $$G(L)= H^2-\left(\omega^2+{\lambda^2\over 4}+\lambda H\right) L^2
+{1\over 4} \lambda^2 L^4.\eqno(5.26)$$
Using $\tilde L = L/2$ one easily sees that
$$ [\tilde L, S_{\pm}]= \pm S_{\pm}, \qquad
[S_+, S_-]= 8\tilde L \left(\omega^2-{\lambda^2\over 4}+\lambda H\right)
-16 \lambda^2 \tilde L^3,\eqno(5.27)$$
which corresponds to an su$_{\Phi}$(2) algebra with
$$ \Phi(J_0(J_0+1))= 4\left(\omega^2-{\lambda^2\over 4}+\lambda H\right)
J_0(J_0+1)-4\lambda^2  (J_0(J_0+1))^2.\eqno(5.28)$$
For $\Phi(x)$ to be an increasing function, the condition
$$ x< {1\over 2 \lambda^2} \left( \omega^2 -{\lambda^2\over 4}
+\lambda H \right)\eqno(5.29) $$
has to be satisfied.

\subsection{ The quadratic Hahn algebra QH(3)}

 The quadratic Hahn algebra QH(3) \cite{GLZ217}
$$ [K_1, K_2]= K_3, \eqno(5.30)$$
$$ [K_2, K_3]= A_2 K_2^2 + C_1 K_1 + D K_2 + G_1,\eqno(5.31)$$
$$ [K_3, K_1]= A_2(K_1 K_2+K_2 K_1) + C_2 K_2 +D K_1 +G_2,\eqno(5.32)$$
can be put in correspondence to an su$_{\Phi}$(2) algebra in the special
case in which $C_1=-1$ and $D=G_2=0$. The equivalence can be seen
\cite{Zhe507} by defining the operators
$$ J_1={J_++J_-\over 2}, \qquad J_2={J_+-J_-\over 2i}.\eqno(5.33)$$
Then the su$_{\Phi}$(2) commutation relations can be written as
$$ [J_0, J_1]= iJ_2, \qquad [J_0, J_2]=-i J_1,\eqno(5.34)$$
$$[J_1, J_2]= {i\over 2} (\Phi(J_0(J_0+1))-\Phi(J_0(J_0-1)).\eqno(5.35)$$
Subsequently one can see that the two algebras are equivalent for
$$ K_1= J_1 + A_2 J_0^2 + G_1, \qquad K_2=J_0, \qquad K_3= -i J_2,
\eqno(5.36)$$
and
$$ \Phi(J_0(J_0+1))= -(2 A_2 G_1 + C_2) J_0(J_0+1) -A_2^2 (J_0(J_0+1))^2.
\eqno(5.37)$$
For $\Phi(x)$ to be an increasing function, the condition
$$ x< -{2 A_2 G_1 + C_2 \over 2 A_2^2} \eqno(5.38)$$
has to be obeyed.

\subsection{ The finite W algebra $\bar{\rm W}_0$}

 It is worth remarking that the finite W algebra $\bar{\rm W}_0$
\cite{Bow945}
$$[U_0, L_0^{\pm}]= \pm L_0^{\pm}, \eqno(5.39)$$
$$ [L_0^+, L_0^-]= (-k(k-1)-2(k+1)h) U_0 +2 (U_0)^3,\eqno(5.40)$$
is also an su$_{\Phi}$(2) algebra with
$$ \Phi(J_0(J_0+1))= \left(-{k(k-1)\over 2}-(k+1)h\right) J_0(J_0+1) +
{1\over 2}  (J_0(J_0+1))^2.\eqno(5.41)$$
For $\Phi(x)$ to be an increasing function, the condition
$$ x > {k(k-1)\over 2} +(k+1)h \eqno(5.42)$$
has to be satisfied.

\bigskip
In all of the above cases the representation theory of the su$_{\Phi}$(2)
algebra immediately follows from eqs. (5.2)--(5.4). In each case the range of
values of the free parameters is limited by the condition that
$\Phi(x)$ has to be an increasing entire function defined for $x\geq -1/4$.
The results of this section are summarized in table 2.

\begin{table}[bth]
\begin{center}
\caption{ Structure functions of generalized deformed su(2) algebras.
For conditions of validity for each of them see the corresponding
subsection of the text. }
\bigskip
\begin{tabular}{|c c p{2.0 in}|}
\hline
\ & $\Phi(J_0(J_0+1))$ & Reference \\
\hline\hline
\romannumeral 1 & $J_0(J_0+1)$ & usual su(2) \\[0.05in]
\romannumeral 2 & $ [J_0]_q [J_0+1]_q $ & su$_q$(2)
\cite{Bie873,Mac4581,SunFu}
\\[0.05in]
\romannumeral 3 & $ 16 \eta J_0(J_0+1) -64 (J_0(J_0+1))^2$ &
           2 identical particles in 2-dim \cite{LM3649}\\[0.05in]
\romannumeral 4 & $ \left(-2H+{\lambda\over 4}\right) J_0(J_0+1) +\lambda
(J_0(J_0+1))^2$ & Kepler system in 2-dim curved space
\cite{Higgs309}\\[0.05in]
\romannumeral 5 & $ 4\left( \omega^2-{\lambda^2\over 4}+\lambda H\right)
J_0(J_0+1)-4 \lambda^2 (J_0(J_0+1))^2 $ & isotropic oscillator in
2-dim curved space \cite{Higgs309} \\[0.05in]
\romannumeral 6 & $ -(2 A_2 G_1 + C_2) J_0(J_0+1) -A_2^2 (J_0(J_0+1))^2 $ &
quadratic Hahn algebra QH(3) \cite{GLZ217}\\[0.05in]
\romannumeral 7 & $ \left(-{k(k-1)\over 2}-(k+1)h\right) J_0(J_0+1) +
{1\over 2}  (J_0(J_0+1))^2 $ & finite W algebra $\bar {\rm W}_0$
\cite{Bow945}\\[0.05in]
\hline
\end{tabular}
\end{center}
\end{table}

\section{ Generalized deformed parafermionic oscillators}

The relation of the above mentioned algebras, and of additional ones,
to generalized deformed parafermions is also worth studying.

 It has been proved \cite{Que245} that any generalized
deformed parafermionic algebra of order $p$ can be written as a generalized
oscillator with structure function
$$ F(x)= x (p+1-x) (\lambda +\mu x+\nu x^2 +\rho x^3 +\sigma x^4 +\ldots),
\eqno(6.1)$$
where $\lambda$, $\mu$, $\nu$, $\rho$, $\sigma$, \dots are real constants
satisfying the conditions
$$ \lambda + \mu x + \nu x^2 + \rho x^3 + \sigma x^4 +\ldots > 0, \qquad
x \in \{ 1,2,\ldots, p\}.\eqno(6.2)$$

\subsection{ The su$_{\Phi}$(2) algebra}

Considering an su$_{\Phi}$(2) algebra \cite{BDK871} with structure function
$$ \Phi(J_0(J_0+1))= A J_0(J_0+1) + B (J_0(J_0+1))^2 + C (J_0(J_0+1))^3,
\eqno(6.3) $$
and making the correspondence
$$ J_+ \to A^{\dag}, \qquad J_-\to A, \qquad J_0\to N,\eqno(6.4)$$
one finds by equating the rhs of the first of eq. (2.4) and the last of eq.
(5.1) that the su$_{\Phi}$(2) algebra is equivalent to a generalized
deformed parafermionic  oscillator of the form
$$F(N)= N (p+1-N) [ -(p^2(p+1)C +p B)+ (p^3 C +(p-1)B) N+
((p^2-p+1)C +B) N^2+ (p-2) C N^3 + C N^4], \eqno(6.5)$$
if the condition
$$ A+ p(p+1) B + p^2 (p+1)^2 C =0 \eqno(6.6)$$
holds. The condition of eq. (6.2) is always satisfied for $B>0$ and $C>0$.

In the special case of $C=0$ one finds that the su$_{\Phi}$(2) algebra
with structure function
$$ \Phi(J_0(J_0+1))= A J_0(J_0+1) + B (J_0(J_0+1))^2\eqno(6.7)$$
is equivalent to a generalized deformed parafermionic oscillator
characterized by
$$ F(N)= B N (p+1-N) (-p+(p-1)N+ N^2),\eqno(6.8)$$
if the condition
$$ A+ p(p+1) B=0\eqno(6.9)$$
is satisfied. The condition of eq. (6.2) is satisfied for $B>0$.

Including higher powers of $J_0(J_0+1)$ in eq. (6.3) results in higher powers
of $N$ in eq. (6.5) and higher powers of $p(p+1)$ in eq. (6.6). If, however,
one sets $B=0$ in eq. (6.7), then eq. (6.8) vanishes, indicating that no
parafermionic oscillator equivalent to the usual su(2) rotator can be
constucted.

\subsection{The finite W algebra $\bar {\rm W}_0$}

 The $\bar {\rm W}_0$ algebra \cite{Bow945}
of eqs (5.39)-(5.40) is equivalent to a generalized
deformed parafermionic algebra with
$$F(N)= N (p+1-N) {1\over 2} (-p+(p-1)N+N^2),\eqno(6.10)$$
provided that the condition
$$ k(k-1)+2(k+1)h=p(p+1) \eqno(6.11)$$
holds. One can easily check that the condition of eq. (6.2) is satisfied
without any further restriction.

\subsection{ Isotropic harmonic oscillator in a 2-dim curved space}

The algebra of the isotropic harmonic oscillator in a 2-dim curved space with
constant curvature $\lambda$ for finite representations can be put in the
form \cite{BDK3700}
$$F(N)= 4 N (p+1-N) \left(\lambda (p+1-N)+\sqrt{\omega^2+\lambda^2/4}\right)
\left( \lambda N +\sqrt{\omega^2+\lambda^2/4}\right),\eqno(6.12)$$
the relevant energy eigenvalues being
$$ E_p = \sqrt{\omega^2 +{\lambda^2\over 4}} (p+1) +{\lambda\over 2}
(p+1)^2,\eqno(6.13)$$
where $\omega$ is the angular frequency of the oscillator. It is clear
that the condition of eq. (6.2) is satisfied without any further
restrictions.

\subsection{ The Kepler problem in a 2-dim curved space}

The algebra of the Kepler problem in a 2-dim curved space with constant
curvature $\lambda$ for finite representations can be put in the form
\cite{BDK3700}
$$ F(N)= N(p+1-N) \left( {4\mu^2\over (p+1)^2} +
\lambda {(p+1-2N)^2\over 4}\right),\eqno(6.14)$$
the corresponding energy eigenvalues being
$$ E_p = -{2\mu^2\over (p+1)^2}+\lambda {p(p+2)\over 8},\eqno(6.15)$$
where $\mu$ is the coefficient of the $-1/r$ term in the Hamiltonian. It is
clear that the restrictions of eq. (6.2) are satisfied automatically.

\subsection{ The Fokas--Lagerstrom potential}

The Fokas--Lagerstrom potential \cite{FL325} is described by the Hamiltonian
$$ H= {1\over 2} (p_x^2+p_y^2)+{x^2\over 2} + {y^2\over 18}.\eqno(6.16)$$
It is therefore an anisotropic oscillator with ratio of frequencies 3:1.
For finite representations it can be seen \cite{BDK3700} that the relevant
algebra can be put in the form
$$ F(N)=16 N (p+1-N) \left(p+{2\over 3}-N\right) \left( p+{4\over 3}-N\right)
\eqno(6.17)$$
for energy eigenvalues $E_p=p+1$, or in the form
$$ F(N)= 16 N (p+1-N) \left( p+{2\over3}-N\right) \left( p+{1\over3}-N\right)
\eqno(6.18)$$
for eigenvalues $E_p=p+2/3$, or in the form
$$ F(N)=16 N (p+1-N) \left(p+{5\over 3}-N\right) \left( p+{4\over 3}-N\right)
\eqno(6.19)$$
for energies $E_p=p+4/3$. In all cases it is clear that the restrictions of
eq. (6.2) are satisfied.

\subsection{ The Smorodinsky--Winternitz potential}

The Smorodinsky--Winternitz potential \cite{WS444} is described by the
Hamiltonian
$$ H={1\over 2} (p_x^2+p_y^2) +k (x^2+y^2) +{c\over x^2},\eqno(6.20)$$
i.e. it is a generalization of the isotropic harmonic oscillator in
two dimensions. For finite representations it can be seen \cite{BDK3700}
that the
relevant algebra takes the form
$$F(N)= 1024 k^2 N (p+1-N) \left( N+{1\over2}\right) \left( p+1+{\sqrt{1+
8 c}\over 2} -N\right)\eqno(6.21)$$
for $c\geq -1/8$ and energy eigenvalues
$$ E_p= \sqrt{8k}\left(p+{5\over 4}+{\sqrt{1+8c}\over 4}\right), \qquad
p=1,2,\ldots.\eqno(6.22)$$
In the special case of $-1/8 \leq c \leq 3/8$ and energy eigenvalues
$$ E_p= \sqrt{8k} \left( p+{5\over 4}-{\sqrt{1+8c}\over 4}\right), \qquad
p=1, 2, \ldots\eqno(6.23)$$
the relevant algebra is
$$F(N)=1024 k^2 N (p+1-N)\left( N+{1\over 2}\right) \left( p+1-
{\sqrt{1+8c}\over 2}-N\right).\eqno(6.24)$$
In both cases the restrictions of eq. (6.2) are satisfied.

\subsection{ Two identical particles in two dimensions}

Using the same procedure as above, the algebra of eq. (5.18)  can be put
in correspondence with a parafermionic oscillator characterized by
$$ F(N)= N(p+1-N) 64 (p+(1-p)N -N^2),\eqno(6.25)$$
if the condition
$$ \eta = 4p(p+1)\eqno(6.26)$$
holds. However, the condition of eq. (6.2) is violated in this case.

\subsection { The quadratic Hahn algebra QH(3)}

For the quadratic Hahn algebra QH(3) of eqs (5.30)-(5.32) one obtains the
parafermionic oscillator with
$$ F(N)= N (p+1-N) A_2^2 (p+(1-p)N-N^2),\eqno(6.27)$$
if the condition
$$ p(p+1) A_2^2+ 2 A_2 G_1 + C_2 =0\eqno(6.28)$$
holds. Again, eq. (6.2) is violated in this case.

\subsection{ The finite W algebra W$_3^{(2)}$}

The finite W algebra W$_3^{(2)}$ \cite{Tjin60,Tjin485,Tjin93,Tjin161}
is characterized by the commutation relations
$$ [H,E]=2E, \qquad [H,F]=-2F, \qquad [E,F]=H^2+C, \eqno(6.29)$$
$$ [C,E]=[C,F]=[C,H]=0.\eqno(6.30)$$
Defining $\tilde H=H/2$ these can be put in the form
$$ [\tilde H, E]=E, \qquad [\tilde H, F]=-F, \qquad [E,F]=4 \tilde H^2 +C,
\eqno(6.31)$$
$$ [C,E]=[C,F]=[C,\tilde H]=0.\eqno(6.32)$$
This algebra is equivalent to a parafermionic oscillator with
$$ F(N)={2\over 3} N(p+1-N) ( 2 p-1+2 N),\eqno(6.33)$$
provided that the condition
$$ C=-{2\over 3} p (2p+1)\eqno(6.34)$$
holds. One can easily see that the condition of eq. (6.2) is satisfied
without any further restriction.

\subsection{ The Holt potential}

The Holt potential \cite{Holt1037}
$$ H = {1\over 2} (p_x^2+p_y^2) +(x^2+4y^2)+{\delta \over x^2}\eqno(6.35) $$
is a generalization of the harmonic oscillator potential with a ratio of
frequencies 2:1. The relevant algebra can be put \cite{BDK3700} in the form
of an oscillator with
$$ F(N)= 2^{23/2}  N (p+1-N) \left( p+1+{\sqrt{1+8\delta}\over 2} -N\right),
\eqno(6.36)$$
where $(1+8\delta)\geq 0$, the relevant energies being given by
$$ E_p= \sqrt{8} \left(p+1+{\sqrt{1+8\delta}\over 4}\right).\eqno(6.37)$$
In this case it is clear that the condition of
eq. (6.2) is always satisfied without any further restrictions.

In the special case $-{1\over 8} \leq \delta \leq {3\over 8}$ one
obtains \cite{BDK3700}
$$ F(N)= 2^{23/2} N (p+1-N) \left(p+1-{\sqrt{1+8\delta}\over 2} -N\right),
\eqno(6.38)$$
the relevant energies being
$$ E_p=\sqrt{8} \left( p+1-{\sqrt{1+8\delta}\over 4}\right).\eqno(6.39)$$
The condition of eq. (6.2) is again satisfied without any further restrictions
within the given range of $\delta$ values.

The deformed oscillator commutation relations in these cases take the
form
$$ [N,A^{\dagger}]=A^{\dagger}, \qquad [N,A]=-A,\eqno(6.40)$$
$$ [A, A^{\dagger}]= 2^{23/2} \left( 3 N^2 -N\left( 4p+1 \pm \sqrt{1+8\delta}
\right) +p^2 \pm {1\over 2} p \sqrt{1+8\delta}\right).\eqno(6.41)$$
It can easily be seen that they are the same as the W$_3^{(2)}$
commutation relations \cite{Tjin60,Tjin485,Tjin93,Tjin161}
with the identifications
$$ F=\sigma A^{\dagger}, \qquad E=\rho A,\qquad C=f(p), \qquad
H=-2N+k(p),\eqno(6.42)$$
where
$$ \rho\sigma = 2^{-19/2} /3,\qquad k(p)={1\over 3} \left( 4p+1\pm
\sqrt{1+8\delta}\right),\eqno(6.43)$$
$$ f(p)= {2\over 9} \left( 14 p^2 +4 p\pm (7p+1)\sqrt{1+8\delta}+1+4\delta
\right).\eqno(6.44)$$
It is thus shown that the Holt potential possesses the W$_3^{(2)}$
symmetry.

The results of this section are summarized in table 3.

\begin{table}[bth]
\begin{center}
\caption{ Structure functions of deformed oscillators. For conditions
of validity and further explanations in the case of the various
generalized deformed parafermionic oscillators see the corresponding
subsection in the text. }
\bigskip
\begin{tabular}{|c c p{2.0 in}|}
\hline
\ & $F(N)$ & Reference \\
\hline\hline
\romannumeral 1 & $N$ & harmonic  oscillator \\[0.05in]
\romannumeral 2 &  ${ {q^N- q^{-N} }  \over {q- q^{-1} } }= [N]_q $ &
$q$-deformed harmonic oscillator \cite{Bie873,Mac4581,SunFu}\\[0.05in]
\romannumeral 3 & $N(p+1-N)$ & parafermionic oscillator \cite{OK82}
 \\[0.05in]
\romannumeral 4 & $ [N]_q [p+1-N]_q$ & $q$-deformed parafermionic oscillator
\cite{FV1019,OKK591} \\[0.05in]
\romannumeral 5 & $ N (p+1-N) (\lambda +\mu N+\nu N^2 +\rho N^3 +\sigma N^4
+\ldots)$ & generalized deformed parafermionic oscillator \cite{Que245}
\\[0.05in]
\romannumeral 6 & $ N (p+1-N) [ -(p^2(p+1)C +p B)+ (p^3 C +(p-1)B) N $ &
3-term su$_{\Phi}$(2) algebra \\
          &$+((p^2-p+1)C +B) N^2+ (p-2) C N^3 + C N^4]$ &
(eq. 6.3)  \\[0.05in]
\romannumeral 7 & $ B N (p+1-N) (-p+(p-1)N+ N^2)$ & 2-term su$_{\Phi}$(2)
algebra (eq.~6.7)\\[0.05in]
\romannumeral 8 & $ N (p+1-N) {1\over 2} (-p+(p-1)N+N^2)$ & finite W algebra
$\bar {\rm W}_0$ \cite{Bow945} \\[0.05in]
\romannumeral 9 & $ 4 N (p+1-N) \left(\lambda (p+1-N)+\sqrt{\omega^2+
\lambda^2/4}\right) $ & isotropic oscillator in 2-dim \\
   & $ \left( \lambda N +\sqrt{\omega^2+\lambda^2/4}\right)$ &
curved space \cite{Higgs309,BDK3700} \\[0.05in]
\romannumeral 10 & $ N(p+1-N) \left( {4\mu^2\over (p+1)^2} +
\lambda {(p+1-2N)^2\over 4}\right)$ & Kepler system in 2-dim curved space
\cite{Higgs309,BDK3700}\\[0.05in]
\romannumeral 11 & $16 N (p+1-N) \left(p+{2\over 3}-N\right) \left( p
+{4\over 3}-N\right)$ & Fokas--Lagerstrom potential \\
             & or  $ 16 N (p+1-N) \left( p+{2\over3}-N\right)
\left( p+{1\over3}-N\right)$ &  \cite{FL325,BDK3700} \\
             & or $ 16 N (p+1-N) \left(p+{5\over 3}-N\right)
\left( p+{4\over 3}-N\right)$ & \\[0.05in]
\romannumeral 12 & $ 1024 k^2 N (p+1-N) \left( N+{1\over2}\right) \left(
p+1\pm {\sqrt{1+
8 c}\over 2} -N\right)$ & Smorodinsky-Winternitz potential
\cite{WS444,BDK3700}
\\[0.05in]
\romannumeral 13 & ${2\over 3} N(p+1-N) ( 2 p-1+2 N)$ & finite W algebra
W$_3^{(2)}$ \cite{Tjin60,Tjin485,Tjin93,Tjin161} \\[0.05in]
\romannumeral 14 & $ 2^{23/2}  N (p+1-N) \left( p+1\pm
{\sqrt{1+8\delta}\over 2} -N\right)$ & Holt potential \cite{Holt1037,BDK3700}
\\[0.05in]
\hline
\end{tabular}
\end{center}
\end{table}

\section{Discussion}

In summary, we have proved that the deformation schemes introduced
in the literature as unifying frameworks for various deformed oscillators
(summarized in table 1)
are equivalent, their unified representation leading to a correspondence
between the deformed oscillator and the N=2 supersymmetric quantum
mechanics (SUSY-QM) scheme.
In addition, we have shown that several physical systems (two identical
particles in two dimensions, isotropic oscillator and Kepler system in a
2-dim curved space) and mathematical structures (quadratic Hahn algebra
QH(3), finite W algebra $\bar {\rm W}_0$) are identified with a
generalized deformed su(2) algebra, the representation theory of which
is known. The results are summarized in table 2.
Furthermore, the generalized deformed parafermionic oscillator
is found to describe several physical systems (isotropic oscillator and
Kepler system in a curved space, Fokas--Lagerstrom,
Smorodinsky--Winternitz and Holt potentials)
and mathematical constructions (generalized deformed su(2) algebras,
finite W algebras $\bar {\rm W}_0$ and W$_3^{(2)}$).
The results are summarized in table 3.
The framework of the
generalized deformed parafermionic oscillator is more general than the
generalized deformed su(2) algebra, since in the rhs of the relevant
basic commutation relation in the former case (first equation in eq. (2.4))
both odd and even
powers are allowed, while in the latter case (eq. (5.1)) only odd powers are
allowed.

The relevance of deformed oscillator algebras, finite W algebras and
qua\-dra\-tic algebras in the study of the symmetries of the anisotropic
quantum harmonic oscillator in two \cite{BDKL003} and three \cite{BDKL218}
dimensions is receiving attention.

\section*{ Acknowledgments}

Support by CEC under contract ERBCHBGCT930467 is gratefully acknowledged
by one of the authors (DB).

\end{document}